\documentclass[floatfix,showpacs,showkeys,amsmath,preprintnumbers,nofootinbib,onecolumn] {revtex4}
\usepackage{graphicx}
\usepackage{colordvi}
\usepackage{amsmath}
\usepackage{amssymb}
\usepackage{latexsym}
\begin{document}

\title{Effect of Vacuum Energy on Evolution of Primordial Black Holes in Einstein Gravity}

\author{Bibekananda Nayak$^{a*}$ and Mubasher Jamil$^{b\dag}$}
\affiliation{{ $^a$Department of Physics, Utkal University, Vanivihar, Bhubaneswar 751004, India}\\
{$^b$Center for Advanced Mathematics and Physics, National University of Sciences and Technology, H-12, Islamabad, Pakistan}\\
Email : $^*$bibeka@iopb.res.in  and  $^{\dag}$mjamil@camp.nust.edu.pk}

\begin{abstract}
\centerline{\bf Abstract} 
We study the evolution of primordail black holes by considering present 
universe is no more matter dominated rather vacuum energy dominated. We also 
consider the accretion of radiation, matter and vacuum energy during 
respective dominance period. In this scenario, we found that radiation 
accretion efficiency should be less than $0.366$ and accretion rate 
is much larger than previous analysis by B. Nayak et al. \cite{ns}. 
Thus here primordial black holes live longer than previous works \cite{ns}. 
Again matter accretion slightly increases the mass and lifetime of 
primordial black holes. However, the vacuum energy accretion is slightly 
complicated one, where accretion is possible only upto a critical time. 
This critical time depends on the values of accretion efficiency and 
formation time. 
If a primordial black hole lives beyond critical time, then its' lifespan 
increases due to  vacuum energy accretion. But for presently evaporating 
primordial black holes, critical time comes much later than their 
evaporating time and  thus vacuum energy could not affect those 
primordial black holes. We again found that the constraints on the 
initial mass fraction of PBH obtained from the $\gamma$-ray background 
limit becomes stronger in the presence of vacuum energy. 
\end{abstract}

\maketitle

%%%%%%%%%%%%%%%%%%%%%%%%%%%%%%%%%%%%%%%%
\section{Introduction}
Black holes which are formed in the early universe are known as Primordial 
Black Holes (PBHs). A comparison of the cosmological density of the 
universe at any time after the Big Bang with the density associated with 
a black hole shows that PBHs would have of order the particle horizon mass. 
PBHs could thus span enormous mass range starting from $10^{-5}gm$ to more 
than $10^{15}gm$. These black holes are formed as a result of initial 
inhomogeneities \cite{zn, bjc}, inflation \cite{kmz, cgl}, phase 
transitions \cite{kp}, bubble collisions \cite{kss, ls}, or the decay of 
cosmic loops \cite{pz}. In 1974 Hawking discovered that the black holes 
emit thermal radiation due to quantum effects \cite{haw}. So the black holes 
get evaporated depending upon their masses.
But in references \cite{ns}, it is shown that evaporation of primordial 
black holes delayed due to accretion of radiation by assuming standard 
picture of Cosmology. Similar kind of works have been done by many other 
authors \cite{mj, fca, jdj}. 

In standard picture of Cosmology \cite{sw}, universe is radiation dominated in early 
period of evolution and is matter dominated now. So universe is undergoing 
a decelerated expansion through out its evolution. But observations of 
distant supernovae and cosmic microwave background anisotropy indicates 
that the present universe is undergoing accelerating expansion \cite{apj}. 
To explain this unwanted observational fact it is thought that present 
universe is dominated by vacuum energy with negative pressure termed as 
dark energy. SN Ia observations also provide the evidence of a decelerated 
universe in the recent past with trasition from decelerated to accelerated 
occuring at redshift $z_{q=0}\sim 0.5$ \cite{mst}. So the vacuum energy 
domination should be started from $z_{q=0}\sim0.5$ 
i.e. $t_{q=0}\sim\frac{1}{2}t_0$.

In this work, we study the evolution of PBH by considering present universe 
is vacuum energy dominated. Here we consider accretion of radiation, matter 
and vacuum energy during respective dominance period and mainly discuss 
how accretion of vacuum energy affect PBH evolution.    

%%%%%%%%%%%%%%%%%%%%%%%%%%%%%%%%%%%%%%%%%%%%%%%%%%
\section{Primordial black holes and Einstein's gravity}

For a spatially flat FRW Universe with scale factor $a$, the first Friedmann equation is
\begin{equation}\label{11}
\Big(\frac{\dot{a}}{a}\Big)^2=\frac{8\pi G}{3}\rho,
\end{equation}
and \\
the total energy conservation equation is
\begin{equation}\label{1a}
\dot \rho+3H(\rho+p)=0,
\end{equation}
where $H=\frac{\dot a}{a}$ is the Hubble parameter, $\rho$ is the
total energy density and $p$ is the total pressure of the background
fluid. Here we assume the equation of state $p=w\rho$ for the cosmic
fluids i.e. radiation ($w=1/3$), matter $w=0$ and vacuum energy $w=-1$. The Universe evolves from an
initial radiation ($t<t_1$) to matter ($t_1<t<t_2$) and finally to vacuum energy
phase ($t>t_2$).\\
Now equation (\ref{1a}) gives \cite{sw, sw2}
\begin{equation}\label{2a}\rho(a)\propto
\begin{cases} a^{-4}  & \, \, (t<t_1),\\
             a^{-3} & \, \,  (t_1<t<t_2),\\
             c_1 & \, \,(t>t_2).\\
\end{cases}\end{equation}
 Using (\ref{2a}) in
(\ref{11}), we obtain \cite{sw, sw2}
\begin{equation}\label{3a}a(t)\propto
\begin{cases} t^{1/2}  & \, \, (t<t_1),\\
             t^{2/3} & \, \,  (t_1<t<t_2),\\
             e^{H_0t} & \, \,(t>t_2), $ where $ H_0=\sqrt{\frac{8\pi G \rho_c}{3}}.\\
\end{cases}\end{equation}

Due to Hawking radiation, the rate at which the PBH mass decreases
is given by
\begin{equation}\label{4a}
\dot M_{\text{evap}}=-\frac{a_H}{256\pi^3}\frac{1}{G^2M^2},
\end{equation}
where $a_H$ is the black body constant.\\
Again PBH mass can be changed by accumulating radiation, matter or vacuum energy at a rate given by
\begin{equation}\label{4b}
\dot M(t)_{\text{acc}}=16\pi G^2 f_iM^2\rho_i,
\end{equation}
where $f$ is the accretion efficiency and $\rho$ is the density. Subscript $i$ indicates radition, matter or vacuum energy.
%%%%%%%%%%%%%%%%%%%%%%%%%%%%%%%%%%%%%%%%%%%%%%%%%%%%%%%%
\section{Different accretion regimes}
\subsection{Accretion of radiation}
When a PBH accretes radiation, the equation
governing this accretion is
\begin{equation}\label{5a}
\dot M(t)_{\text{acc}}=16\pi G^2 f_{rad}M^2\rho_r.
\end{equation}
Making use of equation (\ref{2a}) i.e. $\rho_r=\rho_r^0\Big(\frac{a}{a_0}\Big)^{-4}$ and equation (\ref{3a}), we get
\begin{equation}\label{5b}
\dot M(t)_{\text{acc}}=16\pi
G^2 f_{rad}\rho_{\text{cr}}\Omega_r^0M^2\Big(\frac{t}{t_1}\Big)^{-2}\Big(\frac{t_1}{t_2}\Big)^{-8/3}e^{-4H_0(t_2-t_0)}.
\end{equation}
On integration, equation (\ref{5b}) gives
\begin{equation}\label{5c}
M(t)={M_i}\Big[ 1+ 16\pi
G^2f_{rad}\rho_{\text{cr}}\Omega_r^0M_i\Big(\frac{1}{t_1}\Big)^{-2}\Big(\frac{t_1}{t_2}\Big)^{-8/3}e^{-4H_0(t_2-t_0)}
\Big(\frac{1}{t}-\frac{1}{t_i}\Big)
\Big]^{-1},
\end{equation}
where $M_i$ is an initial mass of PBH at time $t_i$. The superscript
0 refers to present value of physical quantities. Eq. (\ref{5c})
determines the evolution of PBH by the accretion of radiation.

Using horizon mass, which varies with time as $M_H(t)=G^{-1}t$, as
initial mass of PBH and inserting numerical values of different quantities like $G=6.67 \times10^{-8}$ dyne-cm$^2$/gm$^2$, $\rho_c=1.1 \times10^{-29}$ gm/cm$^3$, $t_1=10^{11}s$, $t_2=0.5 \times t_0$ with $t_0=4.42 \times 10^{17}s$ and $\Omega_r^0=10^{-5}$, we get
\begin{equation}\label{5d}
M(t)=M_i\Big[ 1+ 2.729f_{rad}\Big(\frac{t_i}{t}-1\Big)\Big]^{-1}.
\end{equation}
For large time $t$, this equation asymptotes to
\begin{equation}\label{5e}
M(t)=M_i[ 1- 2.729f_{rad}]^{-1},
\end{equation}
which gives for accretion to be effective $f<\frac{1}{2.729} \approx 0.366$. 

The variation of accreting mass with time is shown in figure-1 for different accretion efficiencies. The figure shows that the mass of the PBH increases with increase in accretion efficiency. For a particular accretion efficiency, the mass of the PBH increases for a small period of time and then it becomes constant.

\begin{figure}[h]
\centering
\includegraphics[scale=0.5]{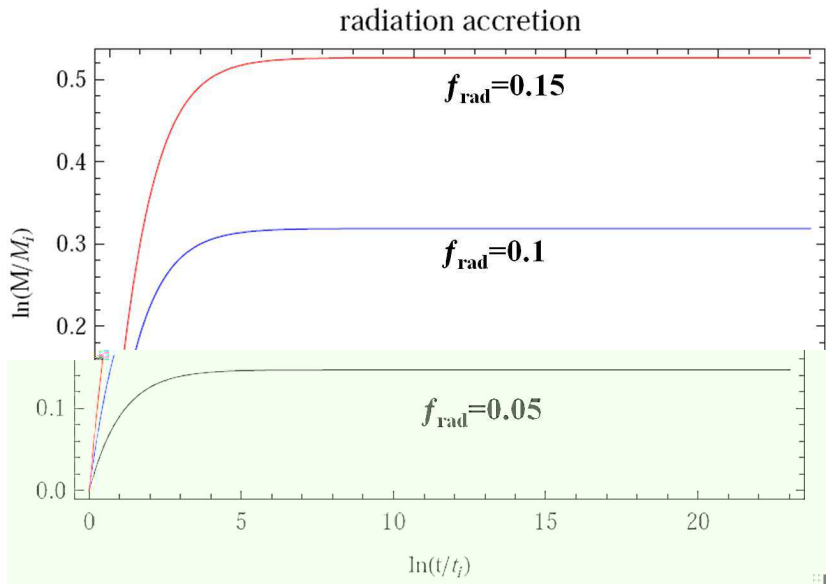}
\caption{Variation of PBH mass for $f_{rad}=0.05,0.1,0.15$}
\label{fig1}
\end{figure}

\subsection{Accretion of matter}
In matter dominated era, PBH mass can increase by absorbing sorrounding matter at a rate
\begin{equation}\label{6a}
\dot M(t)_{\text{acc}}=16\pi G^2 f_{mat}M^2\rho_m.
\end{equation}
In terms of dimensionless density parameters equation (\ref{6a}) can be written by using equations (\ref{2a}) and equation (\ref{3a}) as
\begin{equation}\label{6b}
\dot M(t)_{\text{acc}}=16\pi
G^2f_{mat}M^2\rho_{\text{cr}}\Omega_m^0\Big(\frac{t}{t_2}\Big)^{-2}e^{-3H_0(t_2-t_0)} .
\end{equation}
Integrating equation (\ref{6b}), we get
\begin{equation}\label{6c}
M(t)=M_i\Big[ 1+16\pi G^2f_{mat}\rho_{\text{cr}}t_2^2 M_i
\Omega_m^0 e^{-3H_0(t_2-t_0)}\Big(\frac{1}{t}-\frac{1}{t_i}\Big)
\Big]^{-1}.
\end{equation}
Taking horizon mass as 
initial mass of PBH and using different numerical values along with $\Omega_m^0=0.04$, one can find
\begin{equation}\label{6d}
M(t)={M_i}\Big[ 1+0.372 f_{mat}\Big(\frac{1}{t}-\frac{1}{t_i}\Big)\Big]^{-1}.
\end{equation}
For large time t, this equation gives
\begin{equation}\label{6e}
M(t)={M_i}[ 1-0.372 f_{mat}]^{-1}.
\end{equation}
which gives $f_{mat}$ can take any value between $0$ and $1$.

The variation of accreting mass for different accretion efficiencies is shown in figure-2. This figure indicates that the mass of the PBH varies in a similar fashion as radiation accretion, but here variation is very small. 
\begin{figure}[h]
\centering
\includegraphics[scale=0.5]{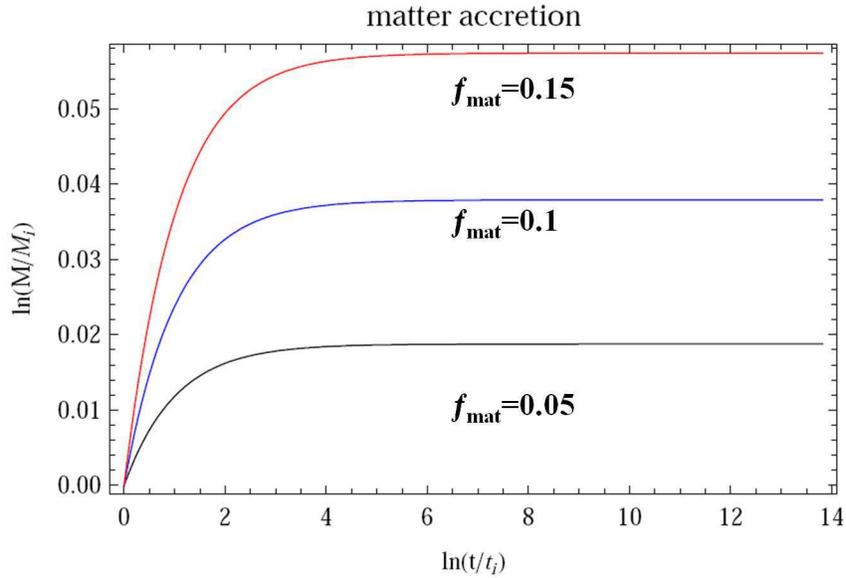}
\caption{Variation of PBH mass for $f_{mat}=0.05,0.1,0.15$}
\label{fig2}
\end{figure}

\subsection{Accretion of vacuum energy}
The presence of vacuum energy affected the mass of the PBH at a rate given by
\begin{equation}\label{7a}
\dot M(t)_{\text{acc}}=16\pi G^2 f_{vac}M^2\rho_{\Lambda}.
\end{equation}
Using equations (\ref{2a}) and (\ref{3a}), we get
\begin{equation}\label{7b}
\dot M(t)_{\text{acc}}=16\pi G^2 f_{vac}M^2\rho_{cr}\Omega_{\Lambda}.
\end{equation}
By integrating equation (\ref{7b}), one can find
\begin{equation}\label{7c}
M(t)=M_i\Big[1+16\pi G^2 f_{vac}M_i\rho_{cr}\Omega_{\Lambda}(t_i-t)\Big]^{-1}.
\end{equation}
Using numerical values of different quantities along with $\Omega_{\Lambda}=0.73$, we get
\begin{equation}\label{7d}
M(t)=M_i[1-2.69 \times10^{-73}f_{vac}M_i(t-t_i)]^{-1}.
\end{equation}
For validity of the above equation, time should be less than a critical value ($t_c$) such that
\begin{equation}\label{7e}
2.69 \times 10^{-73} f_{vac}M_i(t_c-t_i)<1,
\end{equation}
which gives 
\begin{equation}\label{7f}
t_c<\frac{3.717 \times 10^{72}}{f_{vac}M_i}+t_i.
\end{equation}

But if a PBH formed before vacuum dominated era, then its' accreting mass becomes
\begin{equation}\label{7g}
M(t)=M(t_2)[1-2.69 \times10^{-73}f_{vac}M(t_2)(t-t_2)]^{-1},
\end{equation}
which leads to
\begin{equation}\label{7h}
t_c<\frac{3.717 \times 10^{72}}{f_{vac}M(t_2)}+t_2.
\end{equation}

The variation of accreting mass for different accretion efficiencies is shown in figure-3. This figure shows that at a particular time  mass of the PBH increases with increase in accretion efficiency.
\begin{figure}[h]
\centering
\includegraphics[scale=0.5]{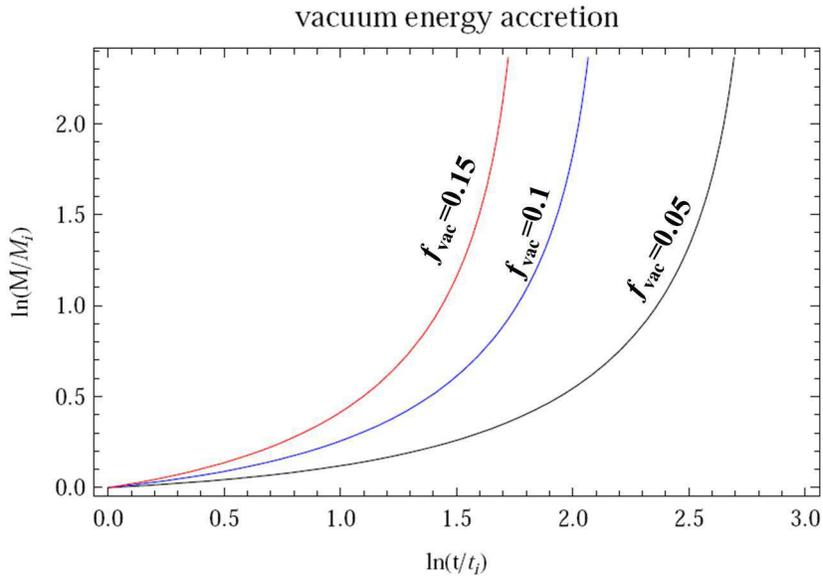}
\caption{Variation of PBH mass for $f_{vac}=0.05$, $0.1$, $0.15$ }
\label{fig3}
\end{figure}
%%%%%%%%%%%%%%%%%%%%%%%%%%%%%%%%%%%%%%%%%%%%%%%%%%%%%%%%%%%%%
\section{PBH evolution in different era}
\subsection{Radiation dominated era}
In radiation dominated era, the mass of the PBH varies accordingly equation
\begin{equation}\label{8a}
\dot M(t)_{\text{PBH}}=16\pi
G^2f_{rad}\rho_{\text{cr}}\Omega_r^0M^2\Big(\frac{t}{t_1}\Big)^{-2}\Big(\frac{t_1}{t_2}\Big)^{-8/3}-\frac{a_H}{256\pi^3}\frac{1}{G^2M^2}.
\end{equation}
This equation can not be solved exactly. But we integrate it by using numerical methods. From our result, we construct the Table-1 for a particular PBH evaporated in the radiation dominated era.
\begin{table}[h]
\begin{tabular}[c]{|c|c|}
\hline
\multicolumn{2}{|c|}{ $t_{i}=10^{-27} s$ }\\
\hline
$f_{rad}$  &  $t_{evap}$    \\
\hline
$0$  &  $3.33\times10^{4}$s   \\
\hline
$0.1$  &  $8.67\times10^{4}$s   \\
\hline
$0.2$  &  $3.55\times10^{5}$s    \\
\hline
$0.3$  &  $5.59\times10^{6}$s    \\
\hline
$0.35$  &  $3.69\times10^{8}$s    \\
\hline
\end{tabular}
\caption{The evaporating times of the PBHs which are created at $t=10^{-27}s$ are displayed for several
radiation accretion efficiencies.}
\end{table}

From the Table-1, we found that the lifetime of PBHs increase with increase in radiation accretion efficiency.

\subsection{Matter dominated era}
Generally PBHs are not formed in matter dominated era, so here we consider about the PBHs which are formed in radiation dominated era. \\
When a PBH passes through matter dominated era, it's mass varies as
\begin{equation}\label{8b}
\dot M(t)_{\text{PBH}}=16\pi
G^2f_{mat}M^2\rho_{\text{cr}}\Omega_m^0\Big(\frac{t}{t_2}\Big)^{-2}-\frac{a_H}{256\pi^3}\frac{1}{G^2M^2}.
\end{equation}
Solving equation (\ref{8b}) along with equation (\ref{8a}) numerically, we construct the Table-2 for a particular PBH evaporating in matter dominated era.
\begin{table}[h]
\begin{tabular}[c]{|c|c|}
\hline
\multicolumn{2}{|c|}{$t_{i}=10^{-25} s$ and $f_{rad}=0.35$}\\
\hline
$f_{mat}$  &  $t_{evap}$    \\
\hline
$0$  &  $3.69\times10^{14}$s   \\
\hline
$0.25$  &  $3.69\times10^{14}$s   \\
\hline
$0.5$  &  $3.69\times10^{14}$s    \\
\hline
$0.75$  &  $3.69\times10^{14}$s    \\
\hline
$1$  &  $3.69\times10^{14}$s   \\
\hline
\end{tabular}
\caption{The evaporating times of the PBHs which are created at $t=10^{-25}s$ are displayed for several matter
accretion efficiencies for a constant radiation accretion.}
\end{table}

Table-2 shows that PBH evolution is not much affected by accretion of matter.
\subsection{Vacuum energy dominated era}
In this era, the mass of the PBH varies at a rate given by
\begin{equation}\label{8c}
\dot M(t)_{\text{PBH}}=16\pi
G^2f_{vac}M^2\rho_{\text{cr}}\Omega_\Lambda^0-\frac{a_H}{256\pi^3}\frac{1}{G^2M^2}.
\end{equation}
But accretion term is only valid upto a critical time $t_c$. After time $t_c$ PBH undergoes only evaporation. i.e.
\begin{equation}\label{8d}
\dot M(t)_{\text{PBH}}=-\frac{a_H}{256\pi^3}\frac{1}{G^2M^2}.
\end{equation}
 
Since PBHs are not formed during vacuum energy dominated era, here we only consider the PBH which are formed in radiation dominated era.\\
Solving equations (\ref{7h}),(\ref{8a}), (\ref{8b}), (\ref{8c}) and (\ref{8d}), 
we construct the Table-3 for a particular PBH evaporating in vacuum energy 
dominated era. It is found from the table that accretion of vacuum energy 
increases the lifespan of PBH. But evaporating time is independent of 
accretion efficiency. Because with increase in accretion efficiency, 
critical time ($t_c$) decreases and $f_{vac}{t_c}$ remains constant. 
Again critical time ($t_c$) for each accretion comes before evaporating time, 
so complete accretion is possible for all accretion efficiencies.  

\begin{table}[h]
\begin{tabular}[c]{|c|c|c|}
\hline
\multicolumn{3}{|c|}{$t_{i}=10^{-10} s$, $f_{rad}=0.35$ and $f_{mat}=1$}\\
\hline
$f_{vac}$  &  $t_c$  &  $t_{evap}$    \\
\hline
$0$  &  -  &  $3.695\times10^{59}$s   \\
\hline
$0.25$  &  $6.668\times10^{43}$s  &  $1.867\times10^{71}$s  \\
\hline
$0.5$  &   $3.334\times10^{43}$s  &  $1.867\times10^{71}$s   \\
\hline
$0.75$  &    $2.223\times10^{43}$s  &  $1.867\times10^{71}$s  \\
\hline
$1$  &   $1.667\times10^{43}$s  &  $1.867\times10^{71}$s \\
\hline
\end{tabular}
\caption{The evaporating times of the PBHs which are created at $t=10^{-10}s$ are displayed for several vacuum energy
accretion efficiencies for a constant radiation and matter accretion.}
\end{table}
%%%%%%%%%%%%%%%%%%%%%%%%%%%%%%%%%%%%%%%%%%%%%%%%%%%%%%%%%%%%%
\section{Constraints on PBH}
Observed astrophysical constraints arise from the presently evaporating PBHs.
So here we discuss about the PBHs whose evaporating time is $t_0$. Solving equations (\ref{8a}), (\ref{8b}), (\ref{8c}) and (\ref{8d}) numerically, we construct the Table-4 for presently evaporating PBHs.
\begin{table}[h]
\begin{tabular}[c]{|c|c|c|}
\hline
\multicolumn{3}{|c|}{$t_{evap}=t_0=4.42 \times10^{17} s$ and $f_{mat}=1$}\\
\hline
$f_{rad}$  &  $M_i$  &  $(M_i)_{vac}$  \\
\hline
$0$  &  $2.367\times10^{15}$g  &  $2.367\times10^{15}$g  \\
\hline
$0.1$  &  $1.721\times10^{15}$g  &  $1.721\times10^{15}$g  \\
\hline
$0.2$  &  $1.075\times10^{15}$g  &  $1.075\times10^{15}$g  \\
\hline
$0.3$  &  $0.429\times10^{15}$g  &  $0.429\times10^{15}$g  \\
\hline
$0.35$  &  $0.106\times10^{15}$g  &  $0.106\times10^{15}$g  \\
\hline
\end{tabular}
\caption{The formation masses of the PBHs which are evaporating now are displayed for several
accretion efficiencies for both cases without vacuum energy accretion ($M_i$) and with vacuum energy accretion ($M_i)_{vac}$.}
\end{table}

It is clear from the Table-4 that vacuum energy accretion does not affect lifetimes of presently evaporating PBHs. Case is same for all other PBHs which are completely evaporated by present time.
\\
\\
Now we calculate the constraint which arises from the present $\gamma$-ray background as follows.\\
The fraction of the Universes' mass going into PBHs at time $t$ is \cite{bjc}
\begin{eqnarray} \label{9a}
\beta(t)=\Big[\frac{\Omega_{PBH}(t)}{\Omega_R}\Big](1+z)^{-1},
\end{eqnarray}
where $\Omega_{PBH}(t)$ is the density parameter associated with PBHs formed at time $t$, $z$ is the redshift associated with time $t$. $\Omega_R$ is the microwave background density having value $10^{-4}$.\\
For $t<t_1$, redshift defination implies, $(1+z)^{-1}=\Big(\frac{t}{t_1}\Big)^{\frac{1}{2}} \Big(\frac{t_1}{t_2}\Big)^{\frac{2}{3}}e^{H_0(t_2-t_0)}$ . \\

Now using $M=G^{-1}t$, we can write the fraction of the Universe going into PBHs' as a function of mass M as
\begin{eqnarray} \label{9b}
\beta(M)=\Big(\frac{M}{M_1}\Big)^{\frac{1}{2}} \Big(\frac{t_1}{t_2}\Big)^{\frac{2}{3}} e^{H_0(t_2-t_0)} \Omega_{PBH}(M) \times 10^4.
\end{eqnarray}
Observations of the cosmolgical deceleration parameter imply $\Omega_{PBH}(M)<1$ over all mass ranges for which PBHs have not evaporated yet. But presently evaporating PBHs($M_*$) generate a $\gamma$-ray background whose most of the energy is appearing at around 100 Mev \cite{idn}. If the fraction of the emitted energy which goes into photons is $\epsilon_{\gamma}$, then the density of the radiation at this energy is expected to be $\Omega_{\gamma}=\epsilon_{\gamma}\Omega_{PBH}(M_*)$.
Since $\epsilon_{\gamma}\sim 0.1$ and the observed $\gamma$-ray background density around $100$ Mev is $\Omega_{\gamma} \sim 10^{-9}$, one gets $\Omega_{PBH}<10^{-8}$ . \\
Now equation (\ref{9b}),therefore, becomes
\begin{eqnarray} \label{9c}
\beta(M_*) < \Big(\frac{M_*}{M_1}\Big)^{\frac{1}{2}} \times \Big(\frac{t_1}{t_2}\Big)^{\frac{2}{3}} e^{H_0(t_2-t_0)} \times 10^{-4}.
\end{eqnarray}
The variation of $\beta(M_*)$ with $f$ drawn from variation of $M_*$ with $f$ is shown in the Table-2. The bound on $\beta(M_*)$ is strengthened as $f$ approaches its maximum value.

\begin{table}[h]
\begin{tabular}[c]{|c|c|c|}
\hline
\multicolumn{3}{|c|}{$t_{evap}=t_0$}\\
\hline
$f_{rad}$  &  $M_*$  &  $\beta(M_*) <$\\
\hline  
$0$  &  $2.367 \times 10^{15}$g  &  $5.23 \times 10^{-26}$\\
\hline
$0.1$  &  $1.721 \times 10^{15}$g  &  $4.46 \times 10^{-26}$\\
\hline
$0.2$  &  $1.075 \times 10^{15}$g  &  $3.525 \times 10^{-26}$\\
\hline
$0.3$  &  $0.429 \times 10^{15}$g  &  $2.227 \times 10^{-26}$\\
\hline
$0.35$  &  $0.106 \times 10^{15}$g  &  $1.107 \times 10^{-26}$\\
\hline
\end{tabular}
\caption{Upper bounds on the initial mass fraction of PBHs that are
evaporating today for various accretion efficiencies $f$.}
\end{table}

But neglecting the presence of vacuum energy, one can find \cite{ns}
\begin{eqnarray} \label{9d}
\beta_0(M_*) < \Big(\frac{M_*}{M_1}\Big)^{\frac{1}{2}} \times \Big(\frac{t_1}{t_0}\Big)^{\frac{2}{3}} \times 10^{-4}.
\end{eqnarray}
Comparison of equations (\ref{9c}) and (\ref{9d}) gives
\begin{eqnarray} \label{9e}
\beta(M_*) \approx 0.364 \times \beta_0(M_*).
\end{eqnarray}
Thus the constraint on the initial mass fraction of PBH obtained from 
the $\gamma-$ray background limit becomes stronger in the presence of vacuum energy.
%%%%%%%%%%%%%%%%%%%%%%%%%%%%%%%%%%%%%%%%%%%%%%%%%%%%%%%%%%%%%
\section{Conclusion}
Here we study the evolution of primordail black holes by considering present 
universe is vacuum energy dominated. In our consideration, we have taken 
that the universe evolves from an initial radiation dominated to matter 
dominated and finally to present vacuum energy phase. We also consider 
that when a PBH passes through radiation domination, matter domination 
and vacuum domination, it accretes radiation, matter and vacuum energy 
respectively. During radiation dominated era, we found that radiation 
accretion efficiency should less than $0.366$ and accretion rate is much 
larger than previous works by B. Nayak et al. \cite{ns}. Thus primordial 
black holes live longer than previous analysis \cite{ns}. In matter dominated 
era, accretion of matter slightly increases the mass and lifetime of 
primordial black holes. However, during vacuum energy dominated era, 
the accumulation of vacuum energy is possible only upto a critical time $t_c$. 
The value of $t_c$ depends on accretion efficiency and formation time. 
If a PBH lives beyond this critical time, then its' lifespan increases due 
to accretion of vacuum energy. But for presently evaporating PBHs, the 
critial time $t_c$ comes much later than their evaporating time. 
So those PBHs are not affected by the presence of vacuum energy. 
We also found that the constraint on the initial mass fraction of PBH obtained 
from the $\gamma-$ray background limit becomes stronger in the 
presence of vacuum energy.

%%%%%%%%%%%%%%%%%%%%%%%%%%%%%%%%%%%%%%%
\section*{Acknowledgements}
%%%%%%%%%%%%%%%%%%%%%%%%%%%%%%%%%%%%%%%
We are thankful to Prof. L. P. Singh, Utkal University, Bhubaneswar, India for his suggestions which greatly help in preparing this manuscript. B. Nayak would like to thank the Council of Scientific and Industrial Research, Government of India, for the award of SRF, F.No. $09/173(0125)/2007-EMR-I$ .

%%%%%%%%%%%%%%%%%%%%%%%%%%%%%%%%%%%%%%%%%%%%%%%%%%%%%%%

%%%%%%%%%%%%%%%%%%%%%%%%%%%%%%%%%%%%%%%%%%%%%%%%%%%%%%%%%%%%

\end{document}